\documentclass[twocolumn,prl,preprintnumbers,superscriptaddress,amsmath,amssymb,english]{revtex4}
\usepackage{amsmath,amssymb,amsfonts,mathrsfs}
\usepackage{amsthm}
\usepackage{CJK}
\usepackage{bm}
\usepackage{babel}
\usepackage{graphicx}
\allowdisplaybreaks
\usepackage{braket}
\usepackage[breaklinks]{hyperref}
\usepackage{color}
\hypersetup{colorlinks=true, linkcolor=blue, citecolor=blue, filecolor=blue, urlcolor=blue}

\begin{document}

\title{Robust Weyl semimetallic phase in face-centered orthogonal $\textrm{C}_{6}$ \\ with helical carbon chains }

\author{Chao Zhang{\color{blue} $^\ddag$ }}
\email{chaozhang@mail.bnu.edu.cn}
\affiliation{School of Materials Science and Engineering, Anhui University of Science and Technology, Huainan 232001, P. R. China}
\affiliation{State Key Laboratory for Environment-friendly Energy Materials, Southwest University of Science and Technology, Mianyang, Sichuan 621010, P. R. China}

\author{Xian-Yong Ding{\color{blue} $^\ddag$}}
\affiliation{School of Materials Science and Engineering, Anhui University of Science and Technology, Huainan 232001, P. R. China}

\author{Li-Yong Gan}
\affiliation{Institute for Structure and Function $\&$ Department of Physics, Chongqing University, Chongqing 400044, P. R. China}

\author{Yu Cao}
\affiliation{School of Materials Science and Engineering, Anhui University of Science and Technology, Huainan 232001, P. R. China}

\author{Bing-Sheng Li}
\affiliation{State Key Laboratory for Environment-friendly Energy Materials, Southwest University of Science and Technology, Mianyang, Sichuan 621010, P. R. China}

\author{Xiaozhi Wu}
\affiliation{Institute for Structure and Function $\&$ Department of Physics, Chongqing University, Chongqing 400044, P. R. China}

\author{Rui Wang}
\email{rcwang@cqu.edu.cn}
\affiliation{Institute for Structure and Function $\&$ Department of Physics, Chongqing University, Chongqing 400044, P. R. China}
\affiliation{Center for Quantum materials and devices, Chongqing University, Chongqing 400044, P. R. China}



\begin{abstract}
The exploration of topological phases in carbon allotropes offers a fascinating avenue to realize topological devices based on carbon materials. Here, using first-principles calculations, we propose a novel metastable carbon allotrope, which possesses exotically helical carbon chains bridged by quadrangle-rings. This unique structure with $sp^{2}$-$sp^{3}$ bonding networks crystallizes in a noncentrosymmetric face-centered orthogonal (fco) lattice with six atoms in a unit cell, thus named fco-$\textrm{C}_{6}$. The considerable stability of fco-$\textrm{C}_{6}$ is confirmed by phonon spectra, elastic constants, and ab initio molecular dynamics simulations. More importantly, fco-$\textrm{C}_{6}$ exhibits extraordinary electronic properties with the minimum number of Weyl points in a time-reversal preserved Weyl system. The symmetry arguments reveal that the Weyl points are guaranteed to lie along the high-symmetry pathes and thus well separated in momentum space, exhibiting the robustness of topologically protected features. We investigate the topological surface states of fco-$\textrm{C}_{6}$ projected on a semi-infinite (010) surface. There are only nontrivial Fermi arcs across the Fermi surface, which facilitates their measurements in experiments and further applications in carbon allotropes.
\end{abstract}



\maketitle

Carbon, one of most versatile elements in the universe, is in favor of forming a rich variety of allotropes. Carbon allotropes  possess  attractive properties, which are determined by the bonding characters in specific crystal structures. Due to the $2s^2 2p^2$ valence electrons, most stable carbon phases usually host dominating $sp^2$, $sp^3$, or $sp^2$-$sp^3$ hybridized bonding characters, such as graphite, diamond, carbon nanotubes \cite{cnt}, fullerenes\cite{fullerene}, graphene \cite{graphene}, and other promising phases \cite{penta-graphene,M-carbon,bctC4,T-carbon,review-carbon}. For various carbon allotropes discovered so far, graphene is of particular importance. The discovery of graphene has attracted intensive topics on two-dimensional (2D) materials. More importantly, graphene significantly promotes the development in fields of topological quantum states \cite{Topo-graphene}. It is well-known that graphene exhibits the unique topological features with massless Dirac fermions \cite{Dirac-graphene}. Due to the extremely weak spin-orbital coupling (SOC) effect, graphene is considered as the prototype of topological semimetals (TSMs). Over the past decade, various TSMs, such as  Dirac semimetals \cite{Dirac-semimetal,Dirac-semimetal-ex,Dirac-semimetal-Nabi}, Weyl semimetals (WSMs) \cite{Topo-semimetal-1,Xu2011,WSMs,WSMs-TaAs,WSM-TaAs-ex,WSM-type2, RuiWang-weyl}, and nodal-line semimetals (NLSM) \cite{NLSM,NLSM-2,NLSM-CaP3} as well as beyond \cite{Bradlynaaf5037,Hourglass,ZhuPhysRevX.6.031003, PhysRevLett.116.186402,TSM,TSM-review}, have been the subject of intense studies. In these semimetallic phases, topologically protected band crossings near the Fermi level give rise to nontrivial fermionic quasiparticles.

Early study of TSMs usually focused on materials including heavy atoms \cite{Dirac-semimetal-Nabi,Topo-semimetal-1,TSM-heavy-1,TSM-heavy-2}, in which the SOC effect plays important roles on the formation of nontrivial band topology in electronic structures. In fact, alternatively, the investigation of topological fermions in materials involving light elements, such as carbon allotropes, is of equal importance. Since the SOC effect in carbon can be negligible, a carbon allotrope can be considered as a spinless system with spin-rotational symmetry. In this case, the time-reversal ($\mathcal{T}$) symmetry satisfies $\mathcal{T}^2=1$ instead of $\mathcal{T}^2=-1$ for a spinful system. Therefore, the carbon allotrope may provide a promising platform to investigate the interplay of nontrivial fermions with crystalline symmetries. Motivated by this exciting avenue, the exploration of TSM phases in carbon allotropes have intensively been performed. As expected, topological fermions were predicted in several three-dimensional (3D) carbon phases \cite{TSM-1,NLSM-bcoC16,TSM-2,TSM-bctC16, Carbonnc, PhysRevLett.120.026402}. While advancements have been very encouraging, the TSM phases predicted to date in carbon allotropes have been limited to NLSMs. The other topological phases, especially for WSMs, have been rarely reported in carbon allotropes, even though there was a work referred to Weyl fermions in a nanostructured carbon phases by manually breaking the inversion ($\mathcal{I}$) symmetry \cite{carbon-weyl}. As the Weyl points (WPs) possess specific chiral charge (i.e., Chern numbers), acting as monopoles in momentum space,  WSMs are of particular interest. Therefore, one would like to explore ideal Weyl fermions in pristine carbon allotropes, which can further accumulate the topological quantum states of mater in carbon-based materials.

\begin{figure*}
\includegraphics[scale=0.55]{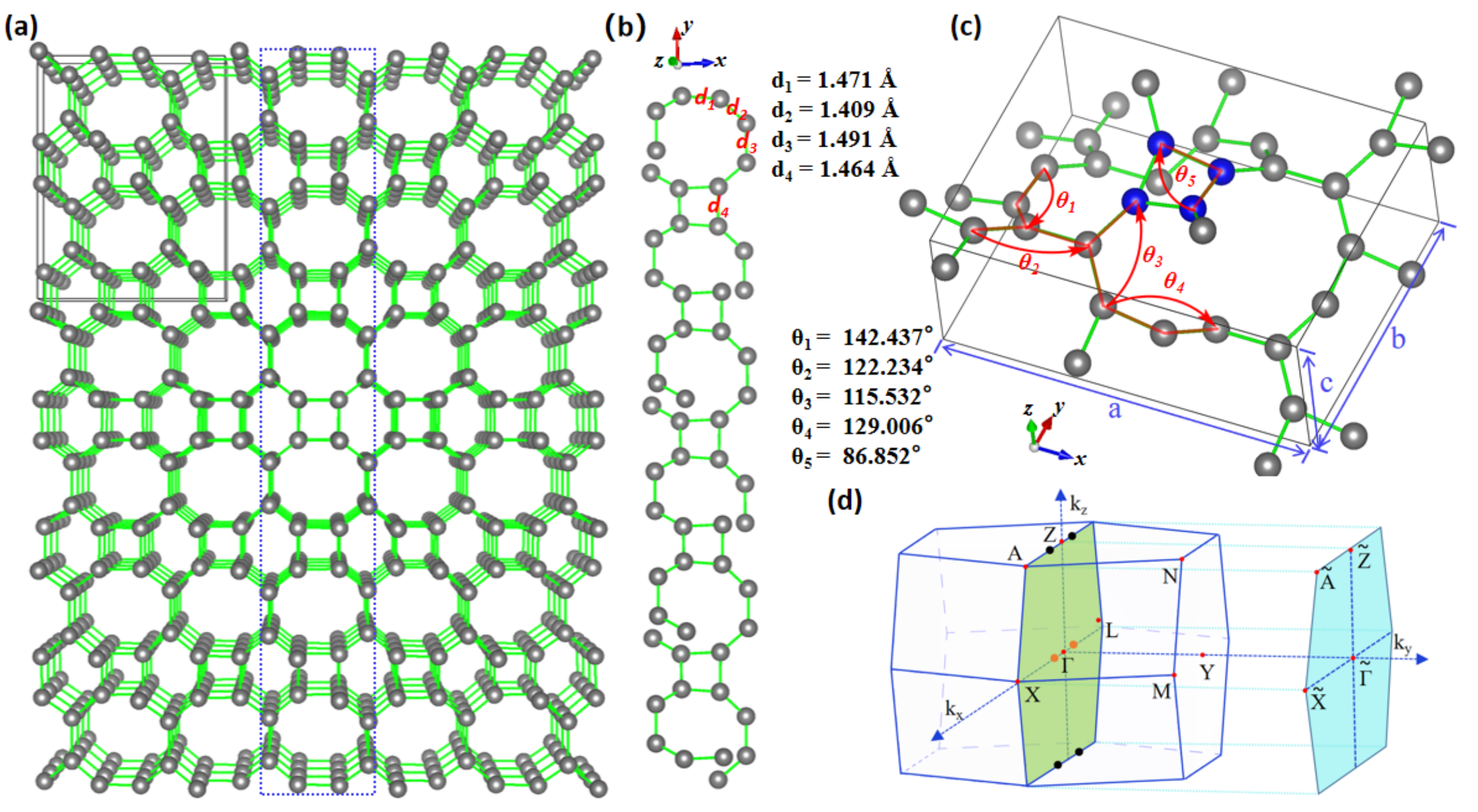}
\caption{Crystal structure and Brillouin zone (BZ) of fco-$\textrm{C}_{6}$. (a) Perspective view of three-dimensional structure. The solid lines depict the conventional unit cell. (b) A helical carbon chain with quadrangle-rings is denoted by the blue dashed-box in (a). (c) The side view of conventional unit cell with the $D_2^7$ space group symmetry (No. 22, $F222$).  The non-equivalent carbon-carbon bond lengths ($d_{1}$, $d_{2}$, $d_{3}$, $d_{4}$) and different angles ($\theta_{1}$, $\theta_{2}$, $\theta_{3}$, $\theta_{4}$, $\theta_{5}$) are also denoted. (d) The fco BZ and the corresponding (010) surface BZ.}
\label{fig1}
\end{figure*}

In this paper, using first-principles calculations, we propose an novel carbon allotrope that possesses topologically protected WSM features. This carbon phase hosts exotically helical carbon chains with quadrangle-rings, exhibiting $sp^{2}$-$sp^{3}$ hybridized bonding characters. It crystallizes in a noncentrosymmetric face-centered orthogonal (fco) structure with a space group $F222$ (No. 22), exhibiting the considerable stability. Furthermore, this carbon allotropic WSM phase only contains two pairs of WPs, i.e., the minimum number of WPs in nonmagnetic WSMs. Due to the extremely weak SOC strength in fco-$\textrm{C}_{6}$, our symmetry and effective model analysis reveal that the WPs are located at high-symmetry pathes of Briblouin zone (BZ).

To depict the structural and electronic properties, we carried out first-principles calculations as implemented in Vienna $ab$ $initio$ simulation package \cite{Kresse1} within the framework of density-functional theory \cite{Kohn1964} (see details in the Supplemental Material \cite{SM}).  The structure of fco-$\textrm{C}_{6}$ is shown in Fig. \ref{fig1}(a). Each carbon atom is surrounded by three neighbors. Due to the non-planar geometries, there are only closed four-membered rings, which are linked by two carbon atoms and  helically arrange along the $y$ direction, forming exotic chiral carbon chains with quadrangle-rings [see Fig. \ref{fig1}(b)]. Hence, the structure of fco-$\textrm{C}_{6}$ can be viewed as formed by linking these chains and stacking along the $x$ direction, exhibiting an unique helical geometry with a $sp^{2}$-$sp^{3}$ hybridized carbon network structure. This structure belongs to the $D_2^7$ space group symmetry (No. 22, $F222$), which lacks the $\mathcal{I}$-symmetry. As illustrated in Fig. \ref{fig1}(c), the optimized lattice constants of fco-$\textrm{C}_{6}$ are $a = 8.835$, $b = 6.816$, and c = 3.351 {\AA}. The six carbon atoms in one unit primitive cell of fco-$\textrm{C}_{6}$ occupy two types of Wyckoff positions as $16k$ (-0.39509, 0.58055, 0.05132) and $8i$ (-0.25 0.83437, 0.25).  There are four non-equivalent carbon-carbon bond lengths and five different angles of fco-$\textrm{C}_{6}$, which are illustrated in Fig. \ref{fig1}. In addition, we also plot the bulk fco BZ and the corresponding (010) surface BZ in Fig. \ref{fig1}(d), in which high-symmetry points are marked.

\begin{figure}
\includegraphics[scale=0.33]{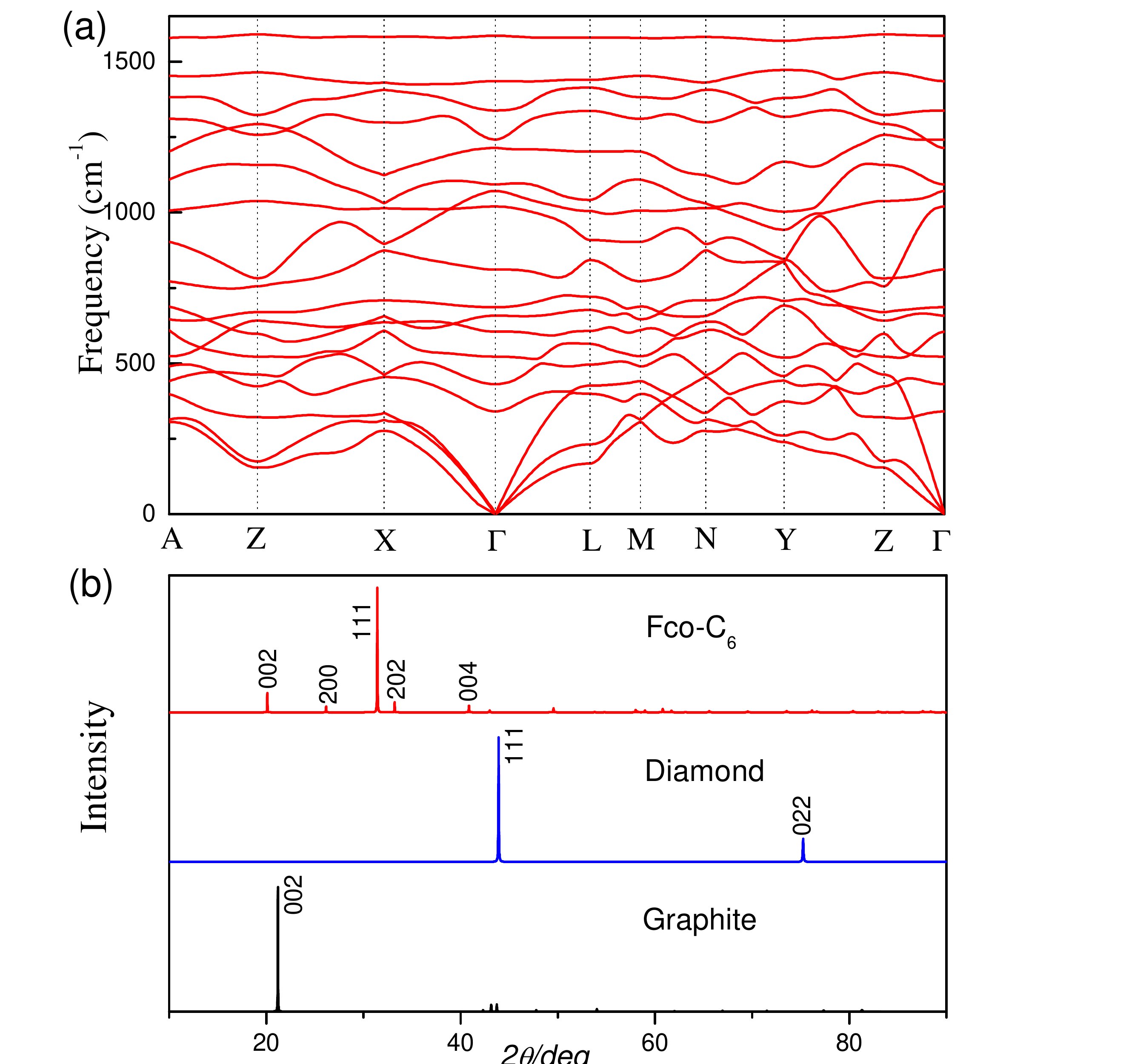}
\caption{(a) The phonon spectra of fco-$\textrm{C}_{6}$. (b) XRD patterns for diamond, graphite and fco-$\textrm{C}_{6}$. }
\label{fig2}
\end{figure}

The calculated cohesive energy $E_{\mathrm{coh}}$ of fco-$\textrm{C}_{6}$ is 7.20 eV/C. It is slightly less than those of graphite and diamond, but is larger than that of T-carbon (6.67 eV/C) \cite{T-carbon}. The more information of comparison with other synthesized carbon phases are provided in Tab. S1 the Supplemental Material \cite{SM}  confirming its energetic stability.  To demonstrate the mechanical stability of fco-$\textrm{C}_{6}$, we calculate the elastic constants. The corresponding calculated results of independent elastic constants $C_{11}$, $C_{12}$, $C_{13}$, $C_{22}$, $C_{23}$, $C_{33}$, $C_{44}$, $C_{55}$, and $C_{66}$ are 661.15, 154.05, 218.93, 128.09, 80.83, 679.62, 139.19, 41.24, and 145.31 GPa, respectively. These values satisfy the Born mechanical stability for the orthogonal crystal \cite{BM-p}, i.e., $C_{11} > 0$; $C_{11}C_{22}$ $>$ $C_{12}^{2}$; $C_{11}C_{22}C_{33}$ + $2C_{12}C_{13}C_{23}$ - $C_{11}C_{23}^{2}$ - $C_{22}C_{13}^{2}$ - $C_{33}C_{12}^{2}$ $>$ 0; $C_{44}$ $>$ 0; $C_{55}$ $>$ 0 and $C_{66}$ $>$ 0, confirming the mechanical stability of fco-$\textrm{C}_{6}$. The calculated $E_{\mathrm{coh}}$ and elastic constants indicate that the structure of fco-$\textrm{C}_{6}$ is very difficult to destroy once it is formed. The dynamical stability of fco-$\textrm{C}_{6}$ can be reflected by the phonon spectra. As shown in Fig. \ref{fig2}(a), there is the absence of the imaginary frequency in the whole BZ. To further investigate the thermodynamical stability of fco-$\textrm{C}_{6}$, the ab initio molecular dynamics simulation was carried out with a 3 $\times$ 3 $\times$ 3 supercell. The structural phase transition did not occur after a relaxation for 5 ps at $T=300$ K. Therefore, fco-$\textrm{C}_{6}$ is an exceptional stable carbon allotrope and may be synthesized in experiments.

The X-ray diffraction (XRD) spectra usually  provide reliable information for experimental observations. The simulated XRD spectra of fco-$\textrm{C}_{6}$ are presented in Fig. \ref{fig2}(b). The graphite and diamond are also supported for comparison. Different from the graphite with the strong peaks of (002) surface at 2$\theta$ = 22.33$^{\circ}$, there are two strong peaks of (002) surface at 20.08$^{\circ}$ and (111) at 31.40$^{\circ}$ in fco-$\textrm{C}_{6}$. Obviously, it can be seen from the XRD spectra that the 2$\theta$ of (002) surface in fco-$\textrm{C}_{6}$ is very close to graphite. This may be because the structure of fco-$\textrm{C}_{6}$ is similar to that of graphite in the Z direction [Fig. \ref{fig1}(c)]. The 2$\theta$ of other weaken peaks are at 26.14$^{\circ}$ of (200) surface, 33.17$^{\circ}$ of (202) furface and 40.82$^{\circ}$ of (004) surface. These characters may be useful to identify the fco-$\textrm{C}_{6}$ in experiments.

\begin{figure}
\includegraphics[scale=0.52]{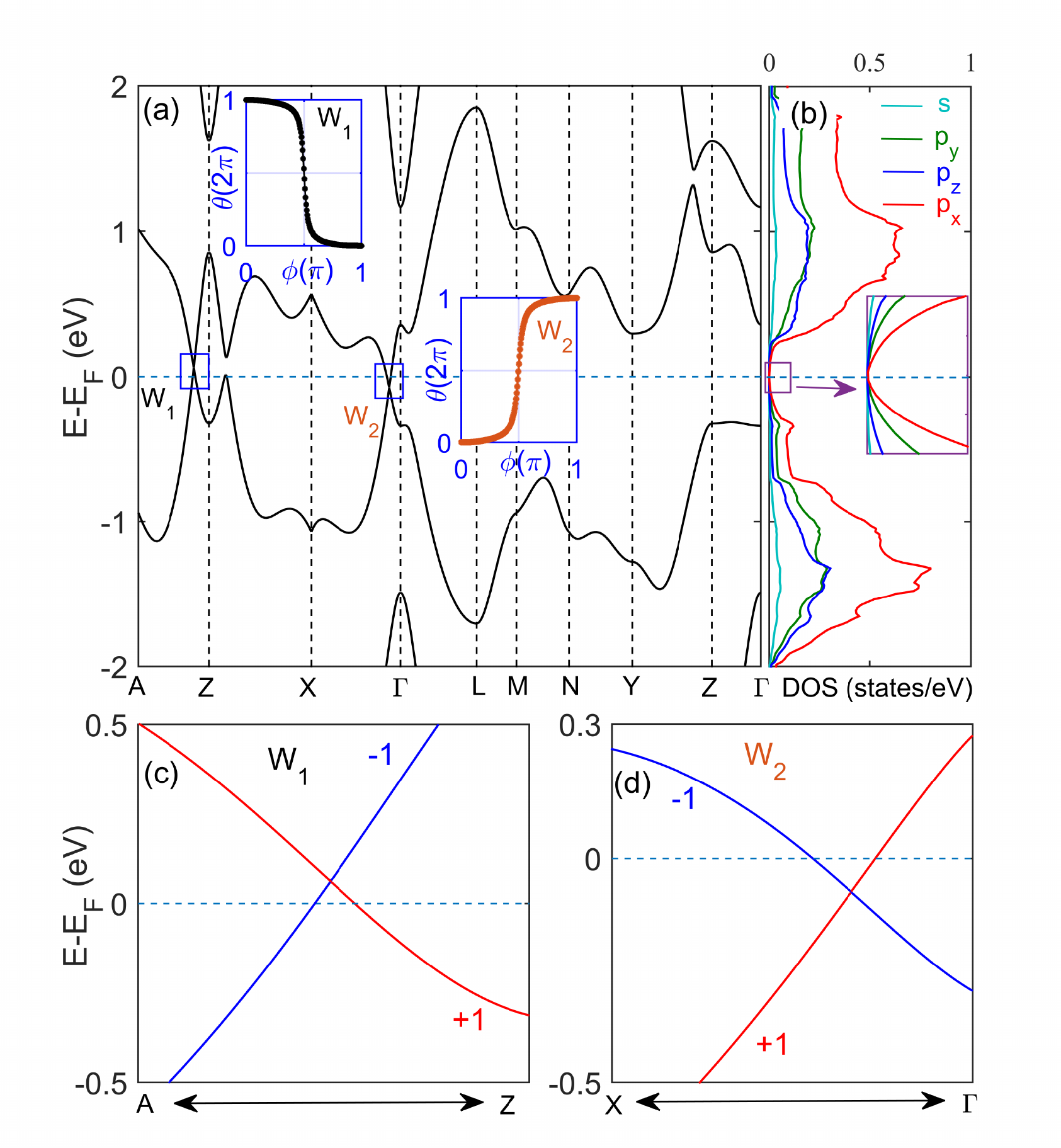}
\caption{(a) The calculated band structure of fco-$\textrm{C}_{6}$ along high-symmetry pathes of fco BZ. The insets show the evolution of Wannier charge centers around the WPs $W_1$ and $W_2$, respectively.  (b) The partial density of states. (c) Enlarged views around the WP $W_1$ along $A$-$Z$; (d) Enlarged views around the WP $W_2$ along  $X$-$\Gamma$. Two crossing bands are represented by opposite eigenvalues $\pm1$ of the little group $C_{2x}$}
\label{fig3}
\end{figure}

In the following, we focus on the electronic properties of fco-$\textrm{C}_{6}$.  The band structures along high-symmetry pathes of the fco BZ are shown in Fig. \ref{fig3}(a). The bands show that there are two nodal points in the $Z$-$A$ and $\Gamma$-$X$ directions, which correspond to the appearance of two bands inverted at the $Z$ and $\Gamma$ points, respectively. The two nodal points are respectively at $\sim$6 meV above and $\sim$7 meV below the Fermi level, indicating that fco-$\textrm{C}_{6}$ is an ideal semimetal. In Fig. \ref{fig3}(b), the partial density of states reveal that the states of $p_x$ orbitals near the Fermi level are much larger than those of other orbitals. However, the contributions of $p_y$ and  $p_z$ orbitals are also visible, confirming $sp^2$-$sp^3$ hybridization of the non-planer carbon network in fco-$\textrm{C}_{6}$. Besides, we also perform the HSE06 calculation, which is often considered as more accurate than the PBE functional, to check the semimetallic feature of fco-$\textrm{C}_{6}$. The comparison between HSE06 and PBE functionals is depicted in Fig. S1 in the Supplemental Material \cite{SM}. The results show that HSE06 functional only slightly shift the band profile, maintaining robust band crossing features.

To further describe the band topology depending on the crystal symmetry, we plot the enlarged views around the nodal points as shown in the Figs. \ref{fig3}(c) and \ref{fig3}(d), respectively.  The crossing bands with linear dispersion along $Z$-$A$ (or $\Gamma$-$X$) are with respect to the twofold rotational symmetry $C_2$ at the $k_x=0$ (or $k_x=\pi/c$) plane. Two crossing bands are represented by opposite eigenvalues $\pm1$ of the little group $C_{2}$. Due to the absence of the $\mathcal{I}$-symmetry in fco-$\textrm{C}_{6}$, the nodal points are the WPs with specific chirality. To determine the chirality of WPs, the Wilson-loop method is employed \cite{WU2017, Yu2011}. As shown in the insets of Fig. \ref{fig3}(a), the evolution of Wannier charge centers shows that the WP (i.e., $W_1$) along $Z$-$A$ possesses the chirality $\mathcal{C}=-1$  and the WP (i.e., $W_2$) along $\Gamma$-$X$ possesses the chirality $\mathcal{C}=+1$. Through carefully screening energy differences between the lowest conduction and highest valence bands, we find that there are totally four WPs over the entire BZ: two are respectively along $Z$-$A$ and $\Gamma$-$X$, and the other two are symmetric with respect to the $\Gamma$ point accompanied by the $\mathcal{T}$-symmetry. As shown in Fig. \ref{fig1}(a), the positions of WPs with $\mathcal{C}=-1$ are at $(\pm k_{x_1}, 0, \pi/c)$ and with $\mathcal{C}=+1$ are at $(\pm k_{x_2}, 0, 0)$, where $k_{x_1}=0.22$ {\AA}$^{-1}$ and $k_{x_2}=0.16$ {\AA}$^{-1}$. The opposite WPs respectively located in the $k_x=0$ and $k_x=\pi/c$ planes are well separated in momentum space, showing the robustness WSM features in fco-$\textrm{C}_{6}$.

Next, we show by symmetry and effective model analysis to understand the symmetry-guaranteed ideal WPs in fco-$\textrm{C}_{6}$. The noncentrosymmetric space $F222$ lacks the $\mathcal{I}$-symmetry and contains three twofold rotational symmetries $C_{2i}$ with the rotational axis along $i$ direction ($i=x$, $y$, or $z$). Generally, we can use a two-band $\mathbf{k}\cdot \mathbf{p}$ model to describe the two crossing bands, and the $2\times 2$ Hamiltonian is
\begin{equation}\label{Hamiltonian}
\mathcal{H}(\mathbf{k})=f_{x}(\mathbf{k})\sigma_{x}+f_{y}(\mathbf{k})\sigma_{y}+f_{z}(\mathbf{k})\sigma_{z},
\end{equation}
where the wavevector $\mathbf{k}=$($k_x$, $k_y$, $k_z$) is referenced to the band inversion points (i. e., $\Gamma$ or $Z$ points), $f_{x,y,z} (\mathbf{k})$ are real functions, and $\sigma_{x,y,z}$ are the Pauli matrices. In Eq. (\ref{Hamiltonian}), we ignore the the kinetic term, which is proportional to the identity matrix, as it is irrelevant to the band crossing points. Since the WPs are present in the $k_z = 0$ or $k_z = \pi/c$ plane, here we consider the $C_{2z}$ and $\mathcal{T}$ symmetries. The product $C_{2z}\mathcal{T}$ denotes a anti-unitary mirror symmetry $\tilde{M}_z$, which constrains Eq. (\ref{Hamiltonian}) as
\begin{equation}\label{MH}
\mathcal{H}(\tilde{M}_z\mathbf{k})=\tilde{M}_z \mathcal{H}(\mathbf{k})\tilde{M}_z^{-1},
\end{equation}
where $\tilde{M}_z\mathbf{k}= (k_{x},k_{y},-k_{z})$. For a spinless system (as SOC effect is negligible for carbon), $\mathcal{T}^2=1$ indicates that the $\mathcal{T}$-operator can be represented as $\mathcal{T}=K$, where $K$ is a complex conjugate operator. Based on the two basis states of $C_2$ operator, the anti-unitary mirror operator $\tilde{M}_z$ can be expressed as $\sigma_z K$. In this case, Eqs. (\ref{Hamiltonian}) and (\ref{MH}) give
\begin{equation}\label{C2z}
\begin{split}
&f_{x}(k_{x},k_{y},k_{z})=-f_{x}(k_{x},k_{y},-k_{z}), \\
&f_{y,z}(k_{x},k_{y},k_{z})= f_{y,z}(k_{x},k_{y},-k_{z}).
\end{split}
\end{equation}
Due to $f_{x,y,z}(\mathbf{k})$ constrained by the periodic condition, Eq. (\ref{C2z}) requires $k_z = {n\pi}/{c}$ ($n\in \mathbb{Z}$) and  $f_{x}(k_x, k_y, {n\pi}/{c})\equiv 0$. This implies that the crossing points can appear at the $k_z = 0$ or $k_z = \pi/c$ plane. In the $k_z = 0$ plane, the $C_{2x}$ commutes with the Hamiltonian $\mathcal{H}(k_{x},k_{y},0)$ as
\begin{equation}\label{Hc2x}
\mathcal{H}(k_x,-k_y,0)=C_{2x} \mathcal{H}(k_x,k_y,0)C_{2x}^{-1}.
\end{equation}
With the basis of $C_2$ rotational eigenvalues, $C_{2x}$ is represented as $\sigma_z$ and we have
\begin{equation}\label{C2x}
\begin{split}
&f_{y}(k_{x},k_{y},0)=-f_{y}(k_{x},-k_{y},0), \\
&f_{z}(k_{x},k_{y},0)= f_{z}(k_{x},-k_{y},0).
\end{split}
\end{equation}
For WPs along $\Gamma$-$X$ (i.e., the $k_x$ axis), $k_y\equiv0$ and Eq. (\ref{C2x}) require $f_{y}(k_{x}, 0, 0)\equiv 0$. The Hamiltonian $\mathcal{H}(\mathbf{k})=f_z(k_x, 0, 0)\sigma_z$ can give zero energy modes, and then two WPs located at $(\pm k_{x_2}, 0, 0)$ are present. By a similar analysis, $C_{2x}$ in the boundary of $k_z = \pi/c$ can also lead to two WPs located at $(\pm k_{x_1}, 0, \pi/c)$. As a result, there are two pairs of WPs protected by the coexistence of the $\mathcal{T}$ and $C_2$ symmetries in the whole BZ, i.e., a minimum number of WPs in a nonmagnetic Weyl system with the $\mathcal{T}$-symmetry. The WPs with opposite chirality are respectively distributed in the $k_z=0$ and $k_z = \pi/c$ planes, and thus guaranteed to be well separated in momentum space.

\begin{figure}
\includegraphics[scale=0.52]{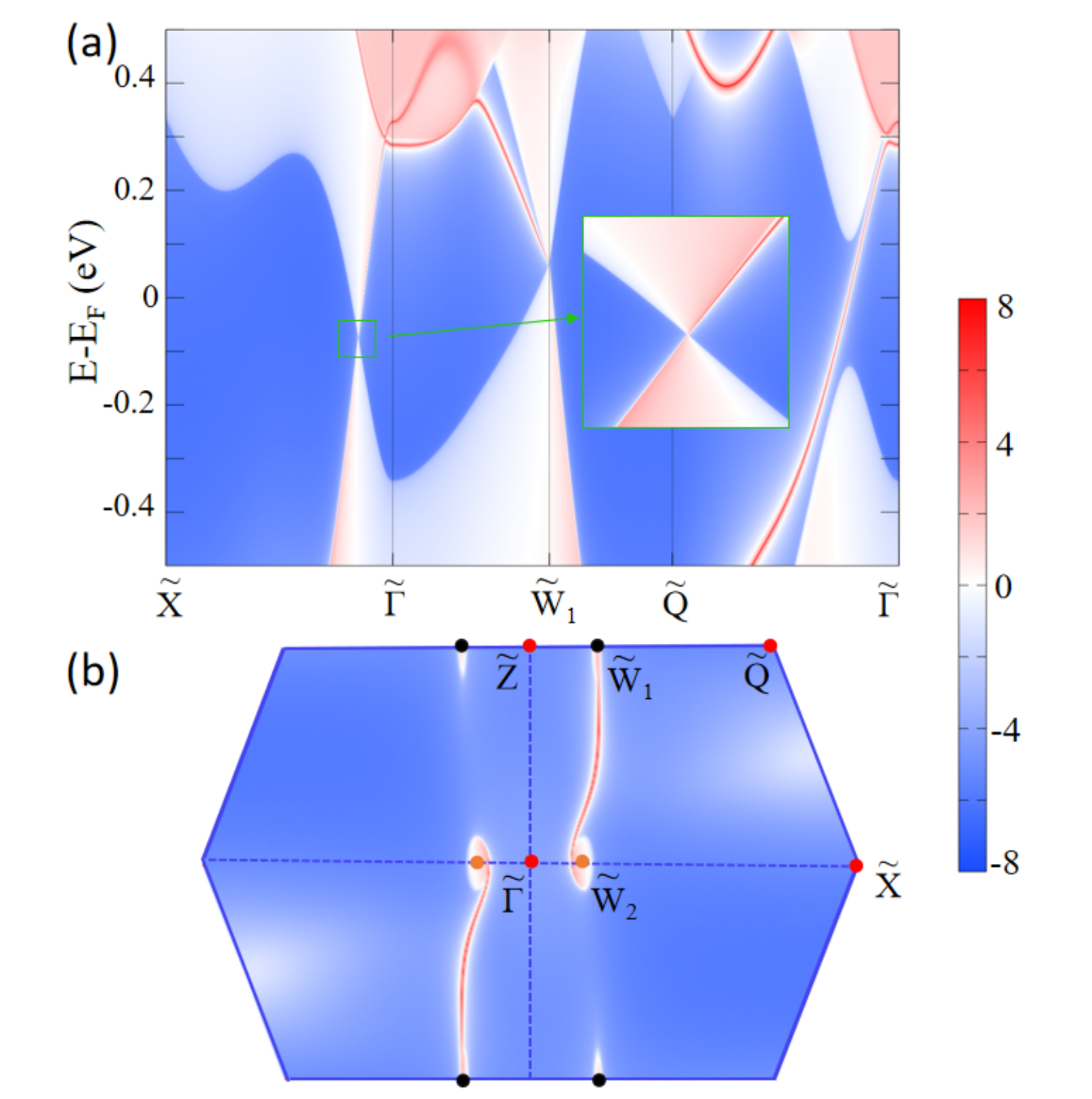}
\caption{ Surface states projected on a semi-infinite (010) surface of fco-$\textrm{C}_{6}$. (a) The calculated LDOS. The inset shows the enlarged view of the WP along $\Gamma$-$X$. (b) The projected Fermi surface. Two Fermi arcs terminated at the projections of WPs with opposite chirality are clearly visible. The orange and black dots represent the projections of WPs with $\mathcal{C}=+1$ and $\mathcal{C}=-1$, respectively.}
\label{fig4}
\end{figure}

The symmetry-guaranteed WPs in fco-$\textrm{C}_{6}$ can possess the unique nontrivial surface states. To proceed this, we constructed a tight-binding (TB) Hamiltonian, which was obtained by projecting from the Bloch states into maximally localized Wannier functions in the WANNIER90 package \cite{Marzari2012,Mostofi2008}. Then, the local density of states (LDOS) and Fermi arcs using the iterative Green's function method  \cite{Sancho1984} \cite{WU2017}. Because of all WPs located at the $k_x-k_z$ plane in momentum space, here we investigate the surface states projected on the semi-infinite (010) surface. In this case, the WPs $W_1$ along $Z$-$A$ and $W_2$ along $\Gamma$-$X$ are respectively projected to $\tilde{Z}$-$\tilde{A}$ and $\tilde{\Gamma}$-$\tilde{X}$ of the (010) surface BZ [see Fig. \ref{fig1}(d)]. The calculated LDOS are plotted in Fig. \ref{fig4}(a). It is found the projections of Dirac cones are clearly visible. The chiral surface states connect the valance and conduction bands, terminated at the projected Dirac cones. The projected Fermi surface of a semi-infinite (010) surface is shown Fig. \ref{fig4}(b). There are two Fermi arcs terminated at the projections of WPs with opposite chirality, corresponding to the two pairs of WPs. Due to the WPs very closed to the Fermi level, we can see that there are nearly only the nontrivial Fermi arcs across the Fermi surface. The trivial bulk states on the (010) surface of fco-$\textrm{C}_{6}$ are absent, which greatly facilitates the observation in experiments.

In summary, using first-principles calculations and symmetry arguments, we show a novel carbon allotrope fco-$\textrm{C}_{6}$ that possesses topologically protected ideal WSM features. This carbon phase with exotically helical carbon chains with quadrangle rings crystallizes a noncentrosymmetric fco structure, exhibiting the $sp^{2}$-$sp^{3}$ hybridized bonding network. The stability analysis shows that fco-$\textrm{C}_{6}$ is an exceptional stable carbon allotrope and prefers to be synthesized in experiments. Importantly, the unique structure and bonding characters of fco-$\textrm{C}_{6}$ leads to fascinating electronic properties with the minimum number of $\mathcal{T}$-preserved WPs.
Due to the extremely weak SOC strength of carbon, the WPs are protected to at high-symmetry pathes of fco BZ by the coexistence of the $\mathcal{T}$ and $C_2$ symmetries. The symmetry-protected WPs are very closed to the Fermi level, leading to nearly ideal nontrivial Fermi arc states.  Our finding not only provides a promising platform for experimentally observing ideal Weyl fermions, but also stimulate further topological applications  in carbon-based materials.

~~~\\
~~~\\

This work was supported by the National Nat-ural Science Foundation of China (NSFC, Grants No. 11974062, and No. 11505003),
the Chongqing National Natural Science Foundation (Grants No. cstc2019jcyj-msxmX0563), the Fundamental Research Funds for the Central Univer-sities of China (Grants No. 2019CDXYWL0029, and No. 2019CDJDWL0005), the Provincial Natural Science Foundation of Anhui No. 1608085QA20, the Lift Engineering of Young Talents of Anhui University of Science and Technology, and the Open Project of State Key Laboratory of Environment-friendly Energy Materials (grants No. 19kfhg03). \\
~~~\\

{\color{blue}{$^\ddag$}}C. Z. and X. Y. D. equally contributed to this work.



\begin{thebibliography}{47}%
\makeatletter
\providecommand \@ifxundefined [1]{%
 \@ifx{#1\undefined}
}%
\providecommand \@ifnum [1]{%
 \ifnum #1\expandafter \@firstoftwo
 \else \expandafter \@secondoftwo
 \fi
}%
\providecommand \@ifx [1]{%
 \ifx #1\expandafter \@firstoftwo
 \else \expandafter \@secondoftwo
 \fi
}%
\providecommand \natexlab [1]{#1}%
\providecommand \enquote  [1]{``#1''}%
\providecommand \bibnamefont  [1]{#1}%
\providecommand \bibfnamefont [1]{#1}%
\providecommand \citenamefont [1]{#1}%
\providecommand \href@noop [0]{\@secondoftwo}%
\providecommand \href [0]{\begingroup \@sanitize@url \@href}%
\providecommand \@href[1]{\@@startlink{#1}\@@href}%
\providecommand \@@href[1]{\endgroup#1\@@endlink}%
\providecommand \@sanitize@url [0]{\catcode `\\12\catcode `\$12\catcode
  `\&12\catcode `\#12\catcode `\^12\catcode `\_12\catcode `\%12\relax}%
\providecommand \@@startlink[1]{}%
\providecommand \@@endlink[0]{}%
\providecommand \url  [0]{\begingroup\@sanitize@url \@url }%
\providecommand \@url [1]{\endgroup\@href {#1}{\urlprefix }}%
\providecommand \urlprefix  [0]{URL }%
\providecommand \Eprint [0]{\href }%
\providecommand \doibase [0]{https://doi.org/}%
\providecommand \selectlanguage [0]{\@gobble}%
\providecommand \bibinfo  [0]{\@secondoftwo}%
\providecommand \bibfield  [0]{\@secondoftwo}%
\providecommand \translation [1]{[#1]}%
\providecommand \BibitemOpen [0]{}%
\providecommand \bibitemStop [0]{}%
\providecommand \bibitemNoStop [0]{.\EOS\space}%
\providecommand \EOS [0]{\spacefactor3000\relax}%
\providecommand \BibitemShut  [1]{\csname bibitem#1\endcsname}%
\let\auto@bib@innerbib\@empty
\bibitem [{\citenamefont {Iijima}(1991)}]{cnt}%
  \BibitemOpen
  \bibfield  {author} {\bibinfo {author} {\bibfnamefont {S.}~\bibnamefont
  {Iijima}},\ }\href {https://doi.org/10.1038/354056a0} {\bibfield  {journal}
  {\bibinfo  {journal} {Nature (london)}\ }\textbf {\bibinfo {volume} {354}},\
  \bibinfo {pages} {56} (\bibinfo {year} {1991})}\BibitemShut {NoStop}%
\bibitem [{\citenamefont {Kroto}\ \emph {et~al.}(1985)\citenamefont {Kroto},
  \citenamefont {Heath}, \citenamefont {O'Brien}, \citenamefont {Curl},\ and\
  \citenamefont {Smalley}}]{fullerene}%
  \BibitemOpen
  \bibfield  {author} {\bibinfo {author} {\bibfnamefont {H.~W.}\ \bibnamefont
  {Kroto}}, \bibinfo {author} {\bibfnamefont {J.~R.}\ \bibnamefont {Heath}},
  \bibinfo {author} {\bibfnamefont {S.~C.}\ \bibnamefont {O'Brien}}, \bibinfo
  {author} {\bibfnamefont {R.~F.}\ \bibnamefont {Curl}},\ and\ \bibinfo
  {author} {\bibfnamefont {R.~E.}\ \bibnamefont {Smalley}},\ }\href
  {https://doi.org/10.1038/318162a0} {\bibfield  {journal} {\bibinfo  {journal}
  {Nature (london)}\ }\textbf {\bibinfo {volume} {318}},\ \bibinfo {pages}
  {162} (\bibinfo {year} {1985})}\BibitemShut {NoStop}%
\bibitem [{\citenamefont {Novoselov}\ \emph {et~al.}(2004)\citenamefont
  {Novoselov}, \citenamefont {Geim}, \citenamefont {Morozov}, \citenamefont
  {Jiang}, \citenamefont {Zhang}, \citenamefont {Dubonos}, \citenamefont
  {Grigorieva},\ and\ \citenamefont {Firsov}}]{graphene}%
  \BibitemOpen
  \bibfield  {author} {\bibinfo {author} {\bibfnamefont {K.~S.}\ \bibnamefont
  {Novoselov}}, \bibinfo {author} {\bibfnamefont {A.~K.}\ \bibnamefont {Geim}},
  \bibinfo {author} {\bibfnamefont {S.~V.}\ \bibnamefont {Morozov}}, \bibinfo
  {author} {\bibfnamefont {D.}~\bibnamefont {Jiang}}, \bibinfo {author}
  {\bibfnamefont {Y.}~\bibnamefont {Zhang}}, \bibinfo {author} {\bibfnamefont
  {S.~V.}\ \bibnamefont {Dubonos}}, \bibinfo {author} {\bibfnamefont {I.~V.}\
  \bibnamefont {Grigorieva}},\ and\ \bibinfo {author} {\bibfnamefont {A.~A.}\
  \bibnamefont {Firsov}},\ }\href {https://doi.org/10.1126/science.1102896}
  {\bibfield  {journal} {\bibinfo  {journal} {Science}\ }\textbf {\bibinfo
  {volume} {306}},\ \bibinfo {pages} {666} (\bibinfo {year}
  {2004})}\BibitemShut {NoStop}%
\bibitem [{\citenamefont {{Zhang}}\ \emph {et~al.}(2015)\citenamefont
  {{Zhang}}, \citenamefont {{Zhou}}, \citenamefont {{Wang}}, \citenamefont
  {{Chen}}, \citenamefont {{Kawazoe}},\ and\ \citenamefont
  {{Jena}}}]{penta-graphene}%
  \BibitemOpen
  \bibfield  {author} {\bibinfo {author} {\bibfnamefont {S.}~\bibnamefont
  {{Zhang}}}, \bibinfo {author} {\bibfnamefont {J.}~\bibnamefont {{Zhou}}},
  \bibinfo {author} {\bibfnamefont {Q.}~\bibnamefont {{Wang}}}, \bibinfo
  {author} {\bibfnamefont {X.}~\bibnamefont {{Chen}}}, \bibinfo {author}
  {\bibfnamefont {Y.}~\bibnamefont {{Kawazoe}}},\ and\ \bibinfo {author}
  {\bibfnamefont {P.}~\bibnamefont {{Jena}}},\ }\href
  {https://doi.org/10.1073/pnas.1416591112} {\bibfield  {journal} {\bibinfo
  {journal} {PNAS.}\ }\textbf {\bibinfo {volume} {112}},\ \bibinfo {pages}
  {2372} (\bibinfo {year} {2015})}\BibitemShut {NoStop}%
\bibitem [{\citenamefont {Li}\ \emph {et~al.}(2009)\citenamefont {Li},
  \citenamefont {Ma}, \citenamefont {Oganov}, \citenamefont {Wang},
  \citenamefont {Wang}, \citenamefont {Xu}, \citenamefont {Cui}, \citenamefont
  {Mao},\ and\ \citenamefont {Zou}}]{M-carbon}%
  \BibitemOpen
  \bibfield  {author} {\bibinfo {author} {\bibfnamefont {Q.}~\bibnamefont
  {Li}}, \bibinfo {author} {\bibfnamefont {Y.}~\bibnamefont {Ma}}, \bibinfo
  {author} {\bibfnamefont {A.~R.}\ \bibnamefont {Oganov}}, \bibinfo {author}
  {\bibfnamefont {H.}~\bibnamefont {Wang}}, \bibinfo {author} {\bibfnamefont
  {H.}~\bibnamefont {Wang}}, \bibinfo {author} {\bibfnamefont {Y.}~\bibnamefont
  {Xu}}, \bibinfo {author} {\bibfnamefont {T.}~\bibnamefont {Cui}}, \bibinfo
  {author} {\bibfnamefont {H.-K.}\ \bibnamefont {Mao}},\ and\ \bibinfo {author}
  {\bibfnamefont {G.}~\bibnamefont {Zou}},\ }\href
  {https://doi.org/10.1103/PhysRevLett.102.175506} {\bibfield  {journal}
  {\bibinfo  {journal} {Phys. Rev. Lett.}\ }\textbf {\bibinfo {volume} {102}},\
  \bibinfo {pages} {175506} (\bibinfo {year} {2009})}\BibitemShut {NoStop}%
\bibitem [{\citenamefont {Umemoto}\ \emph {et~al.}(2010)\citenamefont
  {Umemoto}, \citenamefont {Wentzcovitch}, \citenamefont {Saito},\ and\
  \citenamefont {Miyake}}]{bctC4}%
  \BibitemOpen
  \bibfield  {author} {\bibinfo {author} {\bibfnamefont {K.}~\bibnamefont
  {Umemoto}}, \bibinfo {author} {\bibfnamefont {R.~M.}\ \bibnamefont
  {Wentzcovitch}}, \bibinfo {author} {\bibfnamefont {S.}~\bibnamefont
  {Saito}},\ and\ \bibinfo {author} {\bibfnamefont {T.}~\bibnamefont
  {Miyake}},\ }\href {https://doi.org/10.1103/PhysRevLett.104.125504}
  {\bibfield  {journal} {\bibinfo  {journal} {Phys. Rev. Lett.}\ }\textbf
  {\bibinfo {volume} {104}},\ \bibinfo {pages} {125504} (\bibinfo {year}
  {2010})}\BibitemShut {NoStop}%
\bibitem [{\citenamefont {Sheng}\ \emph {et~al.}(2011)\citenamefont {Sheng},
  \citenamefont {Yan}, \citenamefont {Ye}, \citenamefont {Zheng},\ and\
  \citenamefont {Su}}]{T-carbon}%
  \BibitemOpen
  \bibfield  {author} {\bibinfo {author} {\bibfnamefont {X.~L.}\ \bibnamefont
  {Sheng}}, \bibinfo {author} {\bibfnamefont {Q.~B.}\ \bibnamefont {Yan}},
  \bibinfo {author} {\bibfnamefont {F.}~\bibnamefont {Ye}}, \bibinfo {author}
  {\bibfnamefont {Q.~R.}\ \bibnamefont {Zheng}},\ and\ \bibinfo {author}
  {\bibfnamefont {G.}~\bibnamefont {Su}},\ }\href
  {https://doi.org/10.1103/PhysRevLett.106.155703} {\bibfield  {journal}
  {\bibinfo  {journal} {Phys. Rev. Lett.}\ }\textbf {\bibinfo {volume} {106}},\
  \bibinfo {pages} {155703} (\bibinfo {year} {2011})}\BibitemShut {NoStop}%
\bibitem [{\citenamefont {Hirsch}(2010)}]{review-carbon}%
  \BibitemOpen
  \bibfield  {author} {\bibinfo {author} {\bibfnamefont {A.}~\bibnamefont
  {Hirsch}},\ }\href {https://doi.org/10.1038/nmat2885} {\bibfield  {journal}
  {\bibinfo  {journal} {Nat. Mater.}\ }\textbf {\bibinfo {volume} {9}},\
  \bibinfo {pages} {868} (\bibinfo {year} {2010})}\BibitemShut {NoStop}%
\bibitem [{\citenamefont {Vafek}\ and\ \citenamefont
  {Vishwanath}(2014)}]{Topo-graphene}%
  \BibitemOpen
  \bibfield  {author} {\bibinfo {author} {\bibfnamefont {O.}~\bibnamefont
  {Vafek}}\ and\ \bibinfo {author} {\bibfnamefont {A.}~\bibnamefont
  {Vishwanath}},\ }\href
  {https://doi.org/10.1146/annurev-conmatphys-031113-133841} {\bibfield
  {journal} {\bibinfo  {journal} {Annu. Rev. Condens. Matter Phys.}\ }\textbf
  {\bibinfo {volume} {5}},\ \bibinfo {pages} {83} (\bibinfo {year}
  {2014})}\BibitemShut {NoStop}%
\bibitem [{\citenamefont {Novoselov}\ \emph {et~al.}(2005)\citenamefont
  {Novoselov}, \citenamefont {Geim}, \citenamefont {Morozov}, \citenamefont
  {Jiang}, \citenamefont {Katsnelson}, \citenamefont {Grigorieva},
  \citenamefont {Dubonos},\ and\ \citenamefont {Firsov}}]{Dirac-graphene}%
  \BibitemOpen
  \bibfield  {author} {\bibinfo {author} {\bibfnamefont {K.~S.}\ \bibnamefont
  {Novoselov}}, \bibinfo {author} {\bibfnamefont {A.~K.}\ \bibnamefont {Geim}},
  \bibinfo {author} {\bibfnamefont {S.~V.}\ \bibnamefont {Morozov}}, \bibinfo
  {author} {\bibfnamefont {D.}~\bibnamefont {Jiang}}, \bibinfo {author}
  {\bibfnamefont {M.~I.}\ \bibnamefont {Katsnelson}}, \bibinfo {author}
  {\bibfnamefont {I.~V.}\ \bibnamefont {Grigorieva}}, \bibinfo {author}
  {\bibfnamefont {S.~V.}\ \bibnamefont {Dubonos}},\ and\ \bibinfo {author}
  {\bibfnamefont {A.~A.}\ \bibnamefont {Firsov}},\ }\href
  {https://doi.org/10.1038/nature04233} {\bibfield  {journal} {\bibinfo
  {journal} {nature}\ }\textbf {\bibinfo {volume} {438}},\ \bibinfo {pages}
  {197} (\bibinfo {year} {2005})}\BibitemShut {NoStop}%
\bibitem [{\citenamefont {Young}\ \emph {et~al.}(2012)\citenamefont {Young},
  \citenamefont {Zaheer}, \citenamefont {Teo}, \citenamefont {Kane},
  \citenamefont {Mele},\ and\ \citenamefont {Rappe}}]{Dirac-semimetal}%
  \BibitemOpen
  \bibfield  {author} {\bibinfo {author} {\bibfnamefont {S.~M.}\ \bibnamefont
  {Young}}, \bibinfo {author} {\bibfnamefont {S.}~\bibnamefont {Zaheer}},
  \bibinfo {author} {\bibfnamefont {J.~C.~Y.}\ \bibnamefont {Teo}}, \bibinfo
  {author} {\bibfnamefont {C.~L.}\ \bibnamefont {Kane}}, \bibinfo {author}
  {\bibfnamefont {E.~J.}\ \bibnamefont {Mele}},\ and\ \bibinfo {author}
  {\bibfnamefont {A.~M.}\ \bibnamefont {Rappe}},\ }\href
  {https://doi.org/10.1103/PhysRevLett.108.140405} {\bibfield  {journal}
  {\bibinfo  {journal} {Phys. Rev. Lett.}\ }\textbf {\bibinfo {volume} {108}},\
  \bibinfo {pages} {140405} (\bibinfo {year} {2012})}\BibitemShut {NoStop}%
\bibitem [{\citenamefont {Borisenko}\ \emph {et~al.}(2014)\citenamefont
  {Borisenko}, \citenamefont {Gibson}, \citenamefont {Evtushinsky},
  \citenamefont {Zabolotnyy}, \citenamefont {B\"uchner},\ and\ \citenamefont
  {Cava}}]{Dirac-semimetal-ex}%
  \BibitemOpen
  \bibfield  {author} {\bibinfo {author} {\bibfnamefont {S.}~\bibnamefont
  {Borisenko}}, \bibinfo {author} {\bibfnamefont {Q.}~\bibnamefont {Gibson}},
  \bibinfo {author} {\bibfnamefont {D.}~\bibnamefont {Evtushinsky}}, \bibinfo
  {author} {\bibfnamefont {V.}~\bibnamefont {Zabolotnyy}}, \bibinfo {author}
  {\bibfnamefont {B.}~\bibnamefont {B\"uchner}},\ and\ \bibinfo {author}
  {\bibfnamefont {R.~J.}\ \bibnamefont {Cava}},\ }\href
  {https://doi.org/10.1103/PhysRevLett.113.027603} {\bibfield  {journal}
  {\bibinfo  {journal} {Phys. Rev. Lett.}\ }\textbf {\bibinfo {volume} {113}},\
  \bibinfo {pages} {027603} (\bibinfo {year} {2014})}\BibitemShut {NoStop}%
\bibitem [{\citenamefont {Liu}\ \emph {et~al.}(2014)\citenamefont {Liu},
  \citenamefont {Zhou}, \citenamefont {Zhang}, \citenamefont {Wang},
  \citenamefont {Weng}, \citenamefont {Prabhakaran}, \citenamefont {Mo},
  \citenamefont {Shen}, \citenamefont {Fang}, \citenamefont {Dai} \emph
  {et~al.}}]{Dirac-semimetal-Nabi}%
  \BibitemOpen
  \bibfield  {author} {\bibinfo {author} {\bibfnamefont {Z.}~\bibnamefont
  {Liu}}, \bibinfo {author} {\bibfnamefont {B.}~\bibnamefont {Zhou}}, \bibinfo
  {author} {\bibfnamefont {Y.}~\bibnamefont {Zhang}}, \bibinfo {author}
  {\bibfnamefont {Z.}~\bibnamefont {Wang}}, \bibinfo {author} {\bibfnamefont
  {H.}~\bibnamefont {Weng}}, \bibinfo {author} {\bibfnamefont {D.}~\bibnamefont
  {Prabhakaran}}, \bibinfo {author} {\bibfnamefont {S.-K.}\ \bibnamefont {Mo}},
  \bibinfo {author} {\bibfnamefont {Z.}~\bibnamefont {Shen}}, \bibinfo {author}
  {\bibfnamefont {Z.}~\bibnamefont {Fang}}, \bibinfo {author} {\bibfnamefont
  {X.}~\bibnamefont {Dai}}, \emph {et~al.},\ }\href
  {https://doi.org/10.1126/science.1245085} {\bibfield  {journal} {\bibinfo
  {journal} {Science}\ }\textbf {\bibinfo {volume} {343}},\ \bibinfo {pages}
  {864} (\bibinfo {year} {2014})}\BibitemShut {NoStop}%
\bibitem [{\citenamefont {Wan}\ \emph {et~al.}(2011)\citenamefont {Wan},
  \citenamefont {Turner}, \citenamefont {Vishwanath},\ and\ \citenamefont
  {Savrasov}}]{Topo-semimetal-1}%
  \BibitemOpen
  \bibfield  {author} {\bibinfo {author} {\bibfnamefont {X.}~\bibnamefont
  {Wan}}, \bibinfo {author} {\bibfnamefont {A.~M.}\ \bibnamefont {Turner}},
  \bibinfo {author} {\bibfnamefont {A.}~\bibnamefont {Vishwanath}},\ and\
  \bibinfo {author} {\bibfnamefont {S.~Y.}\ \bibnamefont {Savrasov}},\ }\href
  {https://doi.org/10.1103/PhysRevB.83.205101} {\bibfield  {journal} {\bibinfo
  {journal} {Phys. Rev. B}\ }\textbf {\bibinfo {volume} {83}},\ \bibinfo
  {pages} {205101} (\bibinfo {year} {2011})}\BibitemShut {NoStop}%
\bibitem [{\citenamefont {Xu}\ \emph {et~al.}(2011)\citenamefont {Xu},
  \citenamefont {Weng}, \citenamefont {Wang}, \citenamefont {Dai},\ and\
  \citenamefont {Fang}}]{Xu2011}%
  \BibitemOpen
  \bibfield  {author} {\bibinfo {author} {\bibfnamefont {G.}~\bibnamefont
  {Xu}}, \bibinfo {author} {\bibfnamefont {H.}~\bibnamefont {Weng}}, \bibinfo
  {author} {\bibfnamefont {Z.}~\bibnamefont {Wang}}, \bibinfo {author}
  {\bibfnamefont {X.}~\bibnamefont {Dai}},\ and\ \bibinfo {author}
  {\bibfnamefont {Z.}~\bibnamefont {Fang}},\ }\href@noop {} {\bibfield
  {journal} {\bibinfo  {journal} {Phys. Rev. Lett.}\ }\textbf {\bibinfo
  {volume} {107}},\ \bibinfo {pages} {186806} (\bibinfo {year}
  {2011})}\BibitemShut {NoStop}%
\bibitem [{\citenamefont {Xu}\ \emph {et~al.}(2015)\citenamefont {Xu},
  \citenamefont {Belopolski}, \citenamefont {Alidoust}, \citenamefont
  {Neupane}, \citenamefont {Bian}, \citenamefont {Zhang}, \citenamefont
  {Sankar}, \citenamefont {Chang}, \citenamefont {Yuan}, \citenamefont {Lee}
  \emph {et~al.}}]{WSMs}%
  \BibitemOpen
  \bibfield  {author} {\bibinfo {author} {\bibfnamefont {S.-Y.}\ \bibnamefont
  {Xu}}, \bibinfo {author} {\bibfnamefont {I.}~\bibnamefont {Belopolski}},
  \bibinfo {author} {\bibfnamefont {N.}~\bibnamefont {Alidoust}}, \bibinfo
  {author} {\bibfnamefont {M.}~\bibnamefont {Neupane}}, \bibinfo {author}
  {\bibfnamefont {G.}~\bibnamefont {Bian}}, \bibinfo {author} {\bibfnamefont
  {C.}~\bibnamefont {Zhang}}, \bibinfo {author} {\bibfnamefont
  {R.}~\bibnamefont {Sankar}}, \bibinfo {author} {\bibfnamefont
  {G.}~\bibnamefont {Chang}}, \bibinfo {author} {\bibfnamefont
  {Z.}~\bibnamefont {Yuan}}, \bibinfo {author} {\bibfnamefont {C.-C.}\
  \bibnamefont {Lee}}, \emph {et~al.},\ }\href
  {https://doi.org/10.1126/science.aaa9297} {\bibfield  {journal} {\bibinfo
  {journal} {Science}\ }\textbf {\bibinfo {volume} {349}},\ \bibinfo {pages}
  {613} (\bibinfo {year} {2015})}\BibitemShut {NoStop}%
\bibitem [{\citenamefont {Yang}\ \emph {et~al.}(2015)\citenamefont {Yang},
  \citenamefont {Liu}, \citenamefont {Sun}, \citenamefont {Peng}, \citenamefont
  {Yang}, \citenamefont {Zhang}, \citenamefont {Zhou}, \citenamefont {Zhang},
  \citenamefont {Guo}, \citenamefont {Rahn} \emph {et~al.}}]{WSMs-TaAs}%
  \BibitemOpen
  \bibfield  {author} {\bibinfo {author} {\bibfnamefont {L.}~\bibnamefont
  {Yang}}, \bibinfo {author} {\bibfnamefont {Z.}~\bibnamefont {Liu}}, \bibinfo
  {author} {\bibfnamefont {Y.}~\bibnamefont {Sun}}, \bibinfo {author}
  {\bibfnamefont {H.}~\bibnamefont {Peng}}, \bibinfo {author} {\bibfnamefont
  {H.}~\bibnamefont {Yang}}, \bibinfo {author} {\bibfnamefont {T.}~\bibnamefont
  {Zhang}}, \bibinfo {author} {\bibfnamefont {B.}~\bibnamefont {Zhou}},
  \bibinfo {author} {\bibfnamefont {Y.}~\bibnamefont {Zhang}}, \bibinfo
  {author} {\bibfnamefont {Y.}~\bibnamefont {Guo}}, \bibinfo {author}
  {\bibfnamefont {M.}~\bibnamefont {Rahn}}, \emph {et~al.},\ }\href
  {https://doi.org/10.1038/nphys3425} {\bibfield  {journal} {\bibinfo
  {journal} {Nat. Phys.}\ }\textbf {\bibinfo {volume} {11}},\ \bibinfo {pages}
  {728} (\bibinfo {year} {2015})}\BibitemShut {NoStop}%
\bibitem [{\citenamefont {Lv}\ \emph {et~al.}(2015)\citenamefont {Lv},
  \citenamefont {Weng}, \citenamefont {Fu}, \citenamefont {Wang}, \citenamefont
  {Miao}, \citenamefont {Ma}, \citenamefont {Richard}, \citenamefont {Huang},
  \citenamefont {Zhao}, \citenamefont {Chen}, \citenamefont {Fang},
  \citenamefont {Dai}, \citenamefont {Qian},\ and\ \citenamefont
  {Ding}}]{WSM-TaAs-ex}%
  \BibitemOpen
  \bibfield  {author} {\bibinfo {author} {\bibfnamefont {B.~Q.}\ \bibnamefont
  {Lv}}, \bibinfo {author} {\bibfnamefont {H.~M.}\ \bibnamefont {Weng}},
  \bibinfo {author} {\bibfnamefont {B.~B.}\ \bibnamefont {Fu}}, \bibinfo
  {author} {\bibfnamefont {X.~P.}\ \bibnamefont {Wang}}, \bibinfo {author}
  {\bibfnamefont {H.}~\bibnamefont {Miao}}, \bibinfo {author} {\bibfnamefont
  {J.}~\bibnamefont {Ma}}, \bibinfo {author} {\bibfnamefont {P.}~\bibnamefont
  {Richard}}, \bibinfo {author} {\bibfnamefont {X.~C.}\ \bibnamefont {Huang}},
  \bibinfo {author} {\bibfnamefont {L.~X.}\ \bibnamefont {Zhao}}, \bibinfo
  {author} {\bibfnamefont {G.~F.}\ \bibnamefont {Chen}}, \bibinfo {author}
  {\bibfnamefont {Z.}~\bibnamefont {Fang}}, \bibinfo {author} {\bibfnamefont
  {X.}~\bibnamefont {Dai}}, \bibinfo {author} {\bibfnamefont {T.}~\bibnamefont
  {Qian}},\ and\ \bibinfo {author} {\bibfnamefont {H.}~\bibnamefont {Ding}},\
  }\href {https://doi.org/10.1103/PhysRevX.5.031013} {\bibfield  {journal}
  {\bibinfo  {journal} {Phys. Rev. X}\ }\textbf {\bibinfo {volume} {5}},\
  \bibinfo {pages} {031013} (\bibinfo {year} {2015})}\BibitemShut {NoStop}%
\bibitem [{\citenamefont {Soluyanov}\ \emph {et~al.}(2015)\citenamefont
  {Soluyanov}, \citenamefont {Gresch}, \citenamefont {Wang}, \citenamefont
  {Wu}, \citenamefont {Troyer}, \citenamefont {Dai},\ and\ \citenamefont
  {Bernevig}}]{WSM-type2}%
  \BibitemOpen
  \bibfield  {author} {\bibinfo {author} {\bibfnamefont {A.~A.}\ \bibnamefont
  {Soluyanov}}, \bibinfo {author} {\bibfnamefont {D.}~\bibnamefont {Gresch}},
  \bibinfo {author} {\bibfnamefont {Z.}~\bibnamefont {Wang}}, \bibinfo {author}
  {\bibfnamefont {Q.}~\bibnamefont {Wu}}, \bibinfo {author} {\bibfnamefont
  {M.}~\bibnamefont {Troyer}}, \bibinfo {author} {\bibfnamefont
  {X.}~\bibnamefont {Dai}},\ and\ \bibinfo {author} {\bibfnamefont {B.~A.}\
  \bibnamefont {Bernevig}},\ }\href {https://doi.org/10.1038/nature15768}
  {\bibfield  {journal} {\bibinfo  {journal} {Nature}\ }\textbf {\bibinfo
  {volume} {527}},\ \bibinfo {pages} {495} (\bibinfo {year}
  {2015})}\BibitemShut {NoStop}%
\bibitem [{\citenamefont {Xia}\ \emph {et~al.}(2019)\citenamefont {Xia},
  \citenamefont {Jin}, \citenamefont {Zhao}, \citenamefont {Chen},
  \citenamefont {Zheng}, \citenamefont {Zhao}, \citenamefont {Wang},\ and\
  \citenamefont {Xu}}]{RuiWang-weyl}%
  \BibitemOpen
  \bibfield  {author} {\bibinfo {author} {\bibfnamefont {B.~W.}\ \bibnamefont
  {Xia}}, \bibinfo {author} {\bibfnamefont {Y.~J.}\ \bibnamefont {Jin}},
  \bibinfo {author} {\bibfnamefont {J.~Z.}\ \bibnamefont {Zhao}}, \bibinfo
  {author} {\bibfnamefont {Z.~J.}\ \bibnamefont {Chen}}, \bibinfo {author}
  {\bibfnamefont {B.~B.}\ \bibnamefont {Zheng}}, \bibinfo {author}
  {\bibfnamefont {Y.~J.}\ \bibnamefont {Zhao}}, \bibinfo {author}
  {\bibfnamefont {R.}~\bibnamefont {Wang}},\ and\ \bibinfo {author}
  {\bibfnamefont {H.}~\bibnamefont {Xu}},\ }\href
  {https://doi.org/10.1103/PhysRevLett.122.057205} {\bibfield  {journal}
  {\bibinfo  {journal} {Phys. Rev. Lett.}\ }\textbf {\bibinfo {volume} {122}},\
  \bibinfo {pages} {057205} (\bibinfo {year} {2019})}\BibitemShut {NoStop}%
\bibitem [{\citenamefont {Fang}\ \emph {et~al.}(2016)\citenamefont {Fang},
  \citenamefont {Weng}, \citenamefont {Dai},\ and\ \citenamefont
  {Fang}}]{NLSM}%
  \BibitemOpen
  \bibfield  {author} {\bibinfo {author} {\bibfnamefont {C.}~\bibnamefont
  {Fang}}, \bibinfo {author} {\bibfnamefont {H.}~\bibnamefont {Weng}}, \bibinfo
  {author} {\bibfnamefont {X.}~\bibnamefont {Dai}},\ and\ \bibinfo {author}
  {\bibfnamefont {Z.}~\bibnamefont {Fang}},\ }\href
  {https://doi.org/10.1088/1674-1056/25/11/117106} {\bibfield  {journal}
  {\bibinfo  {journal} {Chin. Phys. B}\ }\textbf {\bibinfo {volume} {25}},\
  \bibinfo {pages} {117106} (\bibinfo {year} {2016})}\BibitemShut {NoStop}%
\bibitem [{\citenamefont {Huang}\ \emph {et~al.}(2016)\citenamefont {Huang},
  \citenamefont {Liu}, \citenamefont {Vanderbilt},\ and\ \citenamefont
  {Duan}}]{NLSM-2}%
  \BibitemOpen
  \bibfield  {author} {\bibinfo {author} {\bibfnamefont {H.}~\bibnamefont
  {Huang}}, \bibinfo {author} {\bibfnamefont {J.}~\bibnamefont {Liu}}, \bibinfo
  {author} {\bibfnamefont {D.}~\bibnamefont {Vanderbilt}},\ and\ \bibinfo
  {author} {\bibfnamefont {W.}~\bibnamefont {Duan}},\ }\href
  {https://doi.org/10.1103/PhysRevB.93.201114} {\bibfield  {journal} {\bibinfo
  {journal} {Phys. Rev. B}\ }\textbf {\bibinfo {volume} {93}},\ \bibinfo
  {pages} {201114} (\bibinfo {year} {2016})}\BibitemShut {NoStop}%
\bibitem [{\citenamefont {Xu}\ \emph {et~al.}(2017)\citenamefont {Xu},
  \citenamefont {Yu}, \citenamefont {Fang}, \citenamefont {Dai},\ and\
  \citenamefont {Weng}}]{NLSM-CaP3}%
  \BibitemOpen
  \bibfield  {author} {\bibinfo {author} {\bibfnamefont {Q.}~\bibnamefont
  {Xu}}, \bibinfo {author} {\bibfnamefont {R.}~\bibnamefont {Yu}}, \bibinfo
  {author} {\bibfnamefont {Z.}~\bibnamefont {Fang}}, \bibinfo {author}
  {\bibfnamefont {X.}~\bibnamefont {Dai}},\ and\ \bibinfo {author}
  {\bibfnamefont {H.}~\bibnamefont {Weng}},\ }\href
  {https://doi.org/10.1103/PhysRevB.95.045136} {\bibfield  {journal} {\bibinfo
  {journal} {Phys. Rev. B}\ }\textbf {\bibinfo {volume} {95}},\ \bibinfo
  {pages} {045136} (\bibinfo {year} {2017})}\BibitemShut {NoStop}%
\bibitem [{\citenamefont {Bradlyn}\ \emph {et~al.}(2016)\citenamefont
  {Bradlyn}, \citenamefont {Cano}, \citenamefont {Wang}, \citenamefont
  {Vergniory}, \citenamefont {Felser}, \citenamefont {Cava},\ and\
  \citenamefont {Bernevig}}]{Bradlynaaf5037}%
  \BibitemOpen
  \bibfield  {author} {\bibinfo {author} {\bibfnamefont {B.}~\bibnamefont
  {Bradlyn}}, \bibinfo {author} {\bibfnamefont {J.}~\bibnamefont {Cano}},
  \bibinfo {author} {\bibfnamefont {Z.}~\bibnamefont {Wang}}, \bibinfo {author}
  {\bibfnamefont {M.~G.}\ \bibnamefont {Vergniory}}, \bibinfo {author}
  {\bibfnamefont {C.}~\bibnamefont {Felser}}, \bibinfo {author} {\bibfnamefont
  {R.~J.}\ \bibnamefont {Cava}},\ and\ \bibinfo {author} {\bibfnamefont
  {B.~A.}\ \bibnamefont {Bernevig}},\ }\href
  {https://doi.org/10.1126/science.aaf5037} {\bibfield  {journal} {\bibinfo
  {journal} {Science}\ }\textbf {\bibinfo {volume} {353}},\ \bibinfo {pages}
  {6299} (\bibinfo {year} {2016})}\BibitemShut {NoStop}%
\bibitem [{\citenamefont {Wang}\ \emph {et~al.}(2017)\citenamefont {Wang},
  \citenamefont {Liu}, \citenamefont {Yu}, \citenamefont {Sheng},\ and\
  \citenamefont {Yang}}]{Hourglass}%
  \BibitemOpen
  \bibfield  {author} {\bibinfo {author} {\bibfnamefont {S.-S.}\ \bibnamefont
  {Wang}}, \bibinfo {author} {\bibfnamefont {Y.}~\bibnamefont {Liu}}, \bibinfo
  {author} {\bibfnamefont {Z.-M.}\ \bibnamefont {Yu}}, \bibinfo {author}
  {\bibfnamefont {X.-L.}\ \bibnamefont {Sheng}},\ and\ \bibinfo {author}
  {\bibfnamefont {S.~A.}\ \bibnamefont {Yang}},\ }\href
  {https://doi.org/10.1038/s41467-017-01986-3} {\bibfield  {journal} {\bibinfo
  {journal} {Nat. Commun.}\ }\textbf {\bibinfo {volume} {8}},\ \bibinfo {pages}
  {1844} (\bibinfo {year} {2017})}\BibitemShut {NoStop}%
\bibitem [{\citenamefont {Zhu}\ \emph {et~al.}(2016)\citenamefont {Zhu},
  \citenamefont {Winkler}, \citenamefont {Wu}, \citenamefont {Li},\ and\
  \citenamefont {Soluyanov}}]{ZhuPhysRevX.6.031003}%
  \BibitemOpen
  \bibfield  {author} {\bibinfo {author} {\bibfnamefont {Z.}~\bibnamefont
  {Zhu}}, \bibinfo {author} {\bibfnamefont {G.~W.}\ \bibnamefont {Winkler}},
  \bibinfo {author} {\bibfnamefont {Q.}~\bibnamefont {Wu}}, \bibinfo {author}
  {\bibfnamefont {J.}~\bibnamefont {Li}},\ and\ \bibinfo {author}
  {\bibfnamefont {A.~A.}\ \bibnamefont {Soluyanov}},\ }\href
  {https://doi.org/10.1103/PhysRevX.6.031003} {\bibfield  {journal} {\bibinfo
  {journal} {Phys. Rev. X}\ }\textbf {\bibinfo {volume} {6}},\ \bibinfo {pages}
  {031003} (\bibinfo {year} {2016})}\BibitemShut {NoStop}%
\bibitem [{\citenamefont {Wieder}\ \emph {et~al.}(2016)\citenamefont {Wieder},
  \citenamefont {Kim}, \citenamefont {Rappe},\ and\ \citenamefont
  {Kane}}]{PhysRevLett.116.186402}%
  \BibitemOpen
  \bibfield  {author} {\bibinfo {author} {\bibfnamefont {B.~J.}\ \bibnamefont
  {Wieder}}, \bibinfo {author} {\bibfnamefont {Y.}~\bibnamefont {Kim}},
  \bibinfo {author} {\bibfnamefont {A.~M.}\ \bibnamefont {Rappe}},\ and\
  \bibinfo {author} {\bibfnamefont {C.~L.}\ \bibnamefont {Kane}},\ }\href
  {https://doi.org/10.1103/PhysRevLett.116.186402} {\bibfield  {journal}
  {\bibinfo  {journal} {Phys. Rev. Lett.}\ }\textbf {\bibinfo {volume} {116}},\
  \bibinfo {pages} {186402} (\bibinfo {year} {2016})}\BibitemShut {NoStop}%
\bibitem [{\citenamefont {Burkov}(2016)}]{TSM}%
  \BibitemOpen
  \bibfield  {author} {\bibinfo {author} {\bibfnamefont {A.}~\bibnamefont
  {Burkov}},\ }\href {https://doi.org/10.1038/nmat4788} {\bibfield  {journal}
  {\bibinfo  {journal} {Nat. Mater.}\ }\textbf {\bibinfo {volume} {15}},\
  \bibinfo {pages} {1145} (\bibinfo {year} {2016})}\BibitemShut {NoStop}%
\bibitem [{\citenamefont {Bernevig}\ \emph {et~al.}(2018)\citenamefont
  {Bernevig}, \citenamefont {Weng}, \citenamefont {Fang},\ and\ \citenamefont
  {Dai}}]{TSM-review}%
  \BibitemOpen
  \bibfield  {author} {\bibinfo {author} {\bibfnamefont {A.}~\bibnamefont
  {Bernevig}}, \bibinfo {author} {\bibfnamefont {H.}~\bibnamefont {Weng}},
  \bibinfo {author} {\bibfnamefont {Z.}~\bibnamefont {Fang}},\ and\ \bibinfo
  {author} {\bibfnamefont {X.}~\bibnamefont {Dai}},\ }\href
  {https://doi.org/10.7566/JPSJ.87.041001} {\bibfield  {journal} {\bibinfo
  {journal} {J. Phys. Soc. Jp.}\ }\textbf {\bibinfo {volume} {87}},\ \bibinfo
  {pages} {041001} (\bibinfo {year} {2018})}\BibitemShut {NoStop}%
\bibitem [{\citenamefont {Takane}\ \emph {et~al.}(2016)\citenamefont {Takane},
  \citenamefont {Wang}, \citenamefont {Souma}, \citenamefont {Nakayama},
  \citenamefont {Trang}, \citenamefont {Sato}, \citenamefont {Takahashi},\ and\
  \citenamefont {Ando}}]{TSM-heavy-1}%
  \BibitemOpen
  \bibfield  {author} {\bibinfo {author} {\bibfnamefont {D.}~\bibnamefont
  {Takane}}, \bibinfo {author} {\bibfnamefont {Z.}~\bibnamefont {Wang}},
  \bibinfo {author} {\bibfnamefont {S.}~\bibnamefont {Souma}}, \bibinfo
  {author} {\bibfnamefont {K.}~\bibnamefont {Nakayama}}, \bibinfo {author}
  {\bibfnamefont {C.~X.}\ \bibnamefont {Trang}}, \bibinfo {author}
  {\bibfnamefont {T.}~\bibnamefont {Sato}}, \bibinfo {author} {\bibfnamefont
  {T.}~\bibnamefont {Takahashi}},\ and\ \bibinfo {author} {\bibfnamefont
  {Y.}~\bibnamefont {Ando}},\ }\href
  {https://doi.org/10.1103/PhysRevB.94.121108} {\bibfield  {journal} {\bibinfo
  {journal} {Phys. Rev. B}\ }\textbf {\bibinfo {volume} {94}},\ \bibinfo
  {pages} {121108} (\bibinfo {year} {2016})}\BibitemShut {NoStop}%
\bibitem [{\citenamefont {Hu}\ \emph {et~al.}(2016)\citenamefont {Hu},
  \citenamefont {Tang}, \citenamefont {Liu}, \citenamefont {Liu}, \citenamefont
  {Zhu}, \citenamefont {Graf}, \citenamefont {Myhro}, \citenamefont {Tran},
  \citenamefont {Lau}, \citenamefont {Wei},\ and\ \citenamefont
  {Mao}}]{TSM-heavy-2}%
  \BibitemOpen
  \bibfield  {author} {\bibinfo {author} {\bibfnamefont {J.}~\bibnamefont
  {Hu}}, \bibinfo {author} {\bibfnamefont {Z.}~\bibnamefont {Tang}}, \bibinfo
  {author} {\bibfnamefont {J.}~\bibnamefont {Liu}}, \bibinfo {author}
  {\bibfnamefont {X.}~\bibnamefont {Liu}}, \bibinfo {author} {\bibfnamefont
  {Y.}~\bibnamefont {Zhu}}, \bibinfo {author} {\bibfnamefont {D.}~\bibnamefont
  {Graf}}, \bibinfo {author} {\bibfnamefont {K.}~\bibnamefont {Myhro}},
  \bibinfo {author} {\bibfnamefont {S.}~\bibnamefont {Tran}}, \bibinfo {author}
  {\bibfnamefont {C.~N.}\ \bibnamefont {Lau}}, \bibinfo {author} {\bibfnamefont
  {J.}~\bibnamefont {Wei}},\ and\ \bibinfo {author} {\bibfnamefont
  {Z.}~\bibnamefont {Mao}},\ }\href
  {https://doi.org/10.1103/PhysRevLett.117.016602} {\bibfield  {journal}
  {\bibinfo  {journal} {Phys. Rev. Lett.}\ }\textbf {\bibinfo {volume} {117}},\
  \bibinfo {pages} {016602} (\bibinfo {year} {2016})}\BibitemShut {NoStop}%
\bibitem [{\citenamefont {Cheng}\ \emph {et~al.}(2016)\citenamefont {Cheng},
  \citenamefont {Du}, \citenamefont {Melnik}, \citenamefont {Kawazoe},\ and\
  \citenamefont {Wen}}]{TSM-1}%
  \BibitemOpen
  \bibfield  {author} {\bibinfo {author} {\bibfnamefont {Y.}~\bibnamefont
  {Cheng}}, \bibinfo {author} {\bibfnamefont {J.}~\bibnamefont {Du}}, \bibinfo
  {author} {\bibfnamefont {R.}~\bibnamefont {Melnik}}, \bibinfo {author}
  {\bibfnamefont {Y.}~\bibnamefont {Kawazoe}},\ and\ \bibinfo {author}
  {\bibfnamefont {B.}~\bibnamefont {Wen}},\ }\href
  {https://doi.org/https://doi.org/10.1016/j.carbon.2015.11.039} {\bibfield
  {journal} {\bibinfo  {journal} {Carbon}\ }\textbf {\bibinfo {volume} {98}},\
  \bibinfo {pages} {468 } (\bibinfo {year} {2016})}\BibitemShut {NoStop}%
\bibitem [{\citenamefont {Wang}\ \emph {et~al.}(2016)\citenamefont {Wang},
  \citenamefont {Weng}, \citenamefont {Nie}, \citenamefont {Fang},
  \citenamefont {Kawazoe},\ and\ \citenamefont {Chen}}]{NLSM-bcoC16}%
  \BibitemOpen
  \bibfield  {author} {\bibinfo {author} {\bibfnamefont {J.-T.}\ \bibnamefont
  {Wang}}, \bibinfo {author} {\bibfnamefont {H.}~\bibnamefont {Weng}}, \bibinfo
  {author} {\bibfnamefont {S.}~\bibnamefont {Nie}}, \bibinfo {author}
  {\bibfnamefont {Z.}~\bibnamefont {Fang}}, \bibinfo {author} {\bibfnamefont
  {Y.}~\bibnamefont {Kawazoe}},\ and\ \bibinfo {author} {\bibfnamefont
  {C.}~\bibnamefont {Chen}},\ }\href
  {https://doi.org/10.1103/PhysRevLett.116.195501} {\bibfield  {journal}
  {\bibinfo  {journal} {Phys. Rev. Lett.}\ }\textbf {\bibinfo {volume} {116}},\
  \bibinfo {pages} {195501} (\bibinfo {year} {2016})}\BibitemShut {NoStop}%
\bibitem [{\citenamefont {Weng}\ \emph {et~al.}(2015)\citenamefont {Weng},
  \citenamefont {Liang}, \citenamefont {Xu}, \citenamefont {Yu}, \citenamefont
  {Fang}, \citenamefont {Dai},\ and\ \citenamefont {Kawazoe}}]{TSM-2}%
  \BibitemOpen
  \bibfield  {author} {\bibinfo {author} {\bibfnamefont {H.}~\bibnamefont
  {Weng}}, \bibinfo {author} {\bibfnamefont {Y.}~\bibnamefont {Liang}},
  \bibinfo {author} {\bibfnamefont {Q.}~\bibnamefont {Xu}}, \bibinfo {author}
  {\bibfnamefont {R.}~\bibnamefont {Yu}}, \bibinfo {author} {\bibfnamefont
  {Z.}~\bibnamefont {Fang}}, \bibinfo {author} {\bibfnamefont {X.}~\bibnamefont
  {Dai}},\ and\ \bibinfo {author} {\bibfnamefont {Y.}~\bibnamefont {Kawazoe}},\
  }\href {https://doi.org/10.1103/PhysRevB.92.045108} {\bibfield  {journal}
  {\bibinfo  {journal} {Phys. Rev. B}\ }\textbf {\bibinfo {volume} {92}},\
  \bibinfo {pages} {045108} (\bibinfo {year} {2015})}\BibitemShut {NoStop}%
\bibitem [{\citenamefont {Cheng}\ \emph {et~al.}(2017)\citenamefont {Cheng},
  \citenamefont {Feng}, \citenamefont {Cao}, \citenamefont {Wen}, \citenamefont
  {Wang}, \citenamefont {Kawazoe},\ and\ \citenamefont {Jena}}]{TSM-bctC16}%
  \BibitemOpen
  \bibfield  {author} {\bibinfo {author} {\bibfnamefont {Y.}~\bibnamefont
  {Cheng}}, \bibinfo {author} {\bibfnamefont {X.}~\bibnamefont {Feng}},
  \bibinfo {author} {\bibfnamefont {X.}~\bibnamefont {Cao}}, \bibinfo {author}
  {\bibfnamefont {B.}~\bibnamefont {Wen}}, \bibinfo {author} {\bibfnamefont
  {Q.}~\bibnamefont {Wang}}, \bibinfo {author} {\bibfnamefont {Y.}~\bibnamefont
  {Kawazoe}},\ and\ \bibinfo {author} {\bibfnamefont {P.}~\bibnamefont
  {Jena}},\ }\href {https://doi.org/10.1002/smll.201602894} {\bibfield
  {journal} {\bibinfo  {journal} {Small}\ }\textbf {\bibinfo {volume} {13}},\
  \bibinfo {pages} {1602894} (\bibinfo {year} {2017})}\BibitemShut {NoStop}%
\bibitem [{\citenamefont {Zhong}\ \emph {et~al.}(2017)\citenamefont {Zhong},
  \citenamefont {Chen}, \citenamefont {Yu}, \citenamefont {Xie}, \citenamefont
  {Wang}, \citenamefont {Yang},\ and\ \citenamefont {Zhang}}]{Carbonnc}%
  \BibitemOpen
  \bibfield  {author} {\bibinfo {author} {\bibfnamefont {C.}~\bibnamefont
  {Zhong}}, \bibinfo {author} {\bibfnamefont {Y.}~\bibnamefont {Chen}},
  \bibinfo {author} {\bibfnamefont {Z.-M.}\ \bibnamefont {Yu}}, \bibinfo
  {author} {\bibfnamefont {Y.}~\bibnamefont {Xie}}, \bibinfo {author}
  {\bibfnamefont {H.}~\bibnamefont {Wang}}, \bibinfo {author} {\bibfnamefont
  {S.~A.}\ \bibnamefont {Yang}},\ and\ \bibinfo {author} {\bibfnamefont
  {S.}~\bibnamefont {Zhang}},\ }\href {https://doi.org/10.1038/ncomms15641}
  {\bibfield  {journal} {\bibinfo  {journal} {Nat. Commun.}\ }\textbf {\bibinfo
  {volume} {8}},\ \bibinfo {pages} {15641} (\bibinfo {year}
  {2017})}\BibitemShut {NoStop}%
\bibitem [{\citenamefont {Wang}\ \emph {et~al.}(2018)\citenamefont {Wang},
  \citenamefont {Nie}, \citenamefont {Weng}, \citenamefont {Kawazoe},\ and\
  \citenamefont {Chen}}]{PhysRevLett.120.026402}%
  \BibitemOpen
  \bibfield  {author} {\bibinfo {author} {\bibfnamefont {J.-T.}\ \bibnamefont
  {Wang}}, \bibinfo {author} {\bibfnamefont {S.}~\bibnamefont {Nie}}, \bibinfo
  {author} {\bibfnamefont {H.}~\bibnamefont {Weng}}, \bibinfo {author}
  {\bibfnamefont {Y.}~\bibnamefont {Kawazoe}},\ and\ \bibinfo {author}
  {\bibfnamefont {C.}~\bibnamefont {Chen}},\ }\href@noop {} {\bibfield
  {journal} {\bibinfo  {journal} {Phys. Rev. Lett.}\ }\textbf {\bibinfo
  {volume} {120}},\ \bibinfo {pages} {026402} (\bibinfo {year}
  {2018})}\BibitemShut {NoStop}%
\bibitem [{\citenamefont {Chen}\ \emph {et~al.}(2015)\citenamefont {Chen},
  \citenamefont {Xie}, \citenamefont {Yang}, \citenamefont {Pan}, \citenamefont
  {Zhang}, \citenamefont {Cohen},\ and\ \citenamefont {Zhang}}]{carbon-weyl}%
  \BibitemOpen
  \bibfield  {author} {\bibinfo {author} {\bibfnamefont {Y.}~\bibnamefont
  {Chen}}, \bibinfo {author} {\bibfnamefont {Y.}~\bibnamefont {Xie}}, \bibinfo
  {author} {\bibfnamefont {S.~A.}\ \bibnamefont {Yang}}, \bibinfo {author}
  {\bibfnamefont {H.}~\bibnamefont {Pan}}, \bibinfo {author} {\bibfnamefont
  {F.}~\bibnamefont {Zhang}}, \bibinfo {author} {\bibfnamefont {M.~L.}\
  \bibnamefont {Cohen}},\ and\ \bibinfo {author} {\bibfnamefont
  {S.}~\bibnamefont {Zhang}},\ }\href
  {https://doi.org/10.1021/acs.nanolett.5b02978} {\bibfield  {journal}
  {\bibinfo  {journal} {Nano. Lett.}\ }\textbf {\bibinfo {volume} {15}},\
  \bibinfo {pages} {6974} (\bibinfo {year} {2015})}\BibitemShut {NoStop}%
\bibitem [{\citenamefont {Kresse}\ and\ \citenamefont
  {Furthm\"uller}(1996)}]{Kresse1}%
  \BibitemOpen
  \bibfield  {author} {\bibinfo {author} {\bibfnamefont {G.}~\bibnamefont
  {Kresse}}\ and\ \bibinfo {author} {\bibfnamefont {J.}~\bibnamefont
  {Furthm\"uller}},\ }\href {https://doi.org/10.1103/PhysRevB.54.11169}
  {\bibfield  {journal} {\bibinfo  {journal} {Phys. Rev. B}\ }\textbf {\bibinfo
  {volume} {54}},\ \bibinfo {pages} {11169} (\bibinfo {year}
  {1996})}\BibitemShut {NoStop}%
\bibitem [{\citenamefont {Hohenberg}\ and\ \citenamefont
  {Kohn}(1964)}]{Kohn1964}%
  \BibitemOpen
  \bibfield  {author} {\bibinfo {author} {\bibfnamefont {P.}~\bibnamefont
  {Hohenberg}}\ and\ \bibinfo {author} {\bibfnamefont {W.}~\bibnamefont
  {Kohn}},\ }\href {https://doi.org/10.1103/PhysRev.136.B864} {\bibfield
  {journal} {\bibinfo  {journal} {Phys. Rev.}\ }\textbf {\bibinfo {volume}
  {136}},\ \bibinfo {pages} {B864} (\bibinfo {year} {1964})}\BibitemShut
  {NoStop}%
\bibitem [{SM()}]{SM}%
  \BibitemOpen
  \href@noop {} {}\bibinfo {note} {Supplemental Material}\BibitemShut {NoStop}%
\bibitem [{\citenamefont {Mouhat}\ and\ \citenamefont {Coudert}(2014)}]{BM-p}%
  \BibitemOpen
  \bibfield  {author} {\bibinfo {author} {\bibfnamefont {F.}~\bibnamefont
  {Mouhat}}\ and\ \bibinfo {author} {\bibfnamefont {F.~X.}\ \bibnamefont
  {Coudert}},\ }\href {https://doi.org/10.1103/PhysRevB.90.224104} {\bibfield
  {journal} {\bibinfo  {journal} {Phys. Rev. B}\ }\textbf {\bibinfo {volume}
  {90}},\ \bibinfo {pages} {224104} (\bibinfo {year} {2014})}\BibitemShut
  {NoStop}%
\bibitem [{\citenamefont {Yu}\ \emph {et~al.}(2017)\citenamefont {Yu},
  \citenamefont {Wu}, \citenamefont {Fang},\ and\ \citenamefont
  {Weng}}]{WU2017}%
  \BibitemOpen
  \bibfield  {author} {\bibinfo {author} {\bibfnamefont {R.}~\bibnamefont
  {Yu}}, \bibinfo {author} {\bibfnamefont {Q.}~\bibnamefont {Wu}}, \bibinfo
  {author} {\bibfnamefont {Z.}~\bibnamefont {Fang}},\ and\ \bibinfo {author}
  {\bibfnamefont {H.}~\bibnamefont {Weng}},\ }\href
  {https://doi.org/10.1103/PhysRevLett.119.036401} {\bibfield  {journal}
  {\bibinfo  {journal} {Phys. Rev. Lett.}\ }\textbf {\bibinfo {volume} {119}},\
  \bibinfo {pages} {036401} (\bibinfo {year} {2017})}\BibitemShut {NoStop}%
\bibitem [{\citenamefont {Yu}\ \emph {et~al.}(2011)\citenamefont {Yu},
  \citenamefont {Qi}, \citenamefont {Bernevig}, \citenamefont {Fang},\ and\
  \citenamefont {Dai}}]{Yu2011}%
  \BibitemOpen
  \bibfield  {author} {\bibinfo {author} {\bibfnamefont {R.}~\bibnamefont
  {Yu}}, \bibinfo {author} {\bibfnamefont {X.~L.}\ \bibnamefont {Qi}}, \bibinfo
  {author} {\bibfnamefont {A.}~\bibnamefont {Bernevig}}, \bibinfo {author}
  {\bibfnamefont {Z.}~\bibnamefont {Fang}},\ and\ \bibinfo {author}
  {\bibfnamefont {X.}~\bibnamefont {Dai}},\ }\href
  {https://doi.org/10.1103/PhysRevB.84.075119} {\bibfield  {journal} {\bibinfo
  {journal} {Phys. Rev. B}\ }\textbf {\bibinfo {volume} {84}},\ \bibinfo
  {pages} {075119} (\bibinfo {year} {2011})}\BibitemShut {NoStop}%
\bibitem [{\citenamefont {Marzari}\ \emph {et~al.}(2012)\citenamefont
  {Marzari}, \citenamefont {Mostofi}, \citenamefont {Yates}, \citenamefont
  {Souza},\ and\ \citenamefont {Vanderbilt}}]{Marzari2012}%
  \BibitemOpen
  \bibfield  {author} {\bibinfo {author} {\bibfnamefont {N.}~\bibnamefont
  {Marzari}}, \bibinfo {author} {\bibfnamefont {A.~A.}\ \bibnamefont
  {Mostofi}}, \bibinfo {author} {\bibfnamefont {J.~R.}\ \bibnamefont {Yates}},
  \bibinfo {author} {\bibfnamefont {I.}~\bibnamefont {Souza}},\ and\ \bibinfo
  {author} {\bibfnamefont {D.}~\bibnamefont {Vanderbilt}},\ }\href
  {https://doi.org/10.1103/RevModPhys.84.1419} {\bibfield  {journal} {\bibinfo
  {journal} {Rev. Mod. Phys.}\ }\textbf {\bibinfo {volume} {84}},\ \bibinfo
  {pages} {1419} (\bibinfo {year} {2012})}\BibitemShut {NoStop}%
\bibitem [{\citenamefont {Mostofi}\ \emph {et~al.}(2008)\citenamefont
  {Mostofi}, \citenamefont {Yates}, \citenamefont {Lee}, \citenamefont {Souza},
  \citenamefont {Vanderbilt},\ and\ \citenamefont {Marzari}}]{Mostofi2008}%
  \BibitemOpen
  \bibfield  {author} {\bibinfo {author} {\bibfnamefont {A.~A.}\ \bibnamefont
  {Mostofi}}, \bibinfo {author} {\bibfnamefont {J.~R.}\ \bibnamefont {Yates}},
  \bibinfo {author} {\bibfnamefont {Y.-S.}\ \bibnamefont {Lee}}, \bibinfo
  {author} {\bibfnamefont {I.}~\bibnamefont {Souza}}, \bibinfo {author}
  {\bibfnamefont {D.}~\bibnamefont {Vanderbilt}},\ and\ \bibinfo {author}
  {\bibfnamefont {N.}~\bibnamefont {Marzari}},\ }\href
  {https://doi.org/https://doi.org/10.1016/j.cpc.2007.11.016} {\bibfield
  {journal} {\bibinfo  {journal} {Comput. Phys. Commun.}\ }\textbf {\bibinfo
  {volume} {178}},\ \bibinfo {pages} {685 } (\bibinfo {year}
  {2008})}\BibitemShut {NoStop}%
\bibitem [{\citenamefont {Sancho}\ \emph {et~al.}(1984)\citenamefont {Sancho},
  \citenamefont {Sancho},\ and\ \citenamefont {Rubio}}]{Sancho1984}%
  \BibitemOpen
  \bibfield  {author} {\bibinfo {author} {\bibfnamefont {M.~L.}\ \bibnamefont
  {Sancho}}, \bibinfo {author} {\bibfnamefont {J.~L.}\ \bibnamefont {Sancho}},\
  and\ \bibinfo {author} {\bibfnamefont {J.}~\bibnamefont {Rubio}},\ }\href
  {http://iopscience.iop.org/0305-4608/14/5/016} {\bibfield  {journal}
  {\bibinfo  {journal} {J. Phys. F.}\ }\textbf {\bibinfo {volume} {14}},\
  \bibinfo {pages} {1205} (\bibinfo {year} {1984})}\BibitemShut {NoStop}%
\end{thebibliography}

%

\end{document}